\documentclass[manuscript,screen]{acmart}
\usepackage[nolist,nohyperlinks]{acronym}
\usepackage[]{breqn}
\usepackage{amsmath}
\usepackage{lipsum} 
\usepackage{array}
\usepackage{varwidth}

\usepackage[skip=0.333\baselineskip]{caption}
\usepackage{subcaption}

\usepackage{amsmath}
\DeclareMathOperator*{\argmax}{arg\,max}

\newcommand{\up}[1]{\textsuperscript{#1}}

\makeatletter \renewcommand\d[1]{\ensuremath{%
  \;\mathrm{d}#1\@ifnextchar\d{\!}{}}}
\makeatother

\newtheorem{theorem}{Theorem}

\theoremstyle{definition}
\newtheorem{proposition}[theorem]{Proposition}

\usepackage{multirow}
\usepackage{lipsum}

\AtBeginDocument{%
  \providecommand\BibTeX{{%
    \normalfont B\kern-0.5em{\scshape i\kern-0.25em b}\kern-0.8em\TeX}}}

\copyrightyear{2023}
\acmYear{2023}
\setcopyright{acmlicensed}\acmConference[RecSys '23]{Seventeenth ACM Conference on Recommender Systems}{September 18--22, 2023}{Singapore, Singapore}
\acmBooktitle{Seventeenth ACM Conference on Recommender Systems (RecSys '23), September 18--22, 2023, Singapore, Singapore}
\acmPrice{15.00}
\acmDOI{10.1145/3604915.3608837}
\acmISBN{979-8-4007-0241-9/23/09}

\DeclareMathOperator\erf{erf}

\def\Cov{{\textrm{Cov}}\,}
\def\Var{{\textrm{Var}}\,}
\def\E{\mathop{\mathbb{E}}\,}
\begin{document}

\title{On the Consistency of Average Embeddings for~Item~Recommendation}

\author{Walid Bendada}
\authornote{Contact author: \href{research@deezer.com}{research@deezer.com}}
\orcid{0009-0001-7510-650X}
\affiliation{
  \institution{Deezer Research \& LAMSADE, Université Paris Dauphine, PSL}
  \city{}
  \country{France}
}

\author{Guillaume Salha-Galvan}
\orcid{0000-0002-2452-1041}
\affiliation{
  \institution{Deezer Research}
  \city{}
  \country{France}
}

\author{Romain Hennequin}
\orcid{0000-0001-8158-5562}
\affiliation{
  \institution{Deezer Research}
  \city{}
  \country{France}
}

\author{Thomas Bouabça}
\orcid{0009-0001-0627-9722}
\affiliation{
  \institution{Deezer Research}
  \city{}
  \country{France}
}

\author{Tristan Cazenave}
 \orcid{0000-0003-4669-9374}
\affiliation{
  \institution{LAMSADE, Université Paris Dauphine, PSL}
  \city{}
  \country{France}
}

\renewcommand{\shortauthors}{W. Bendada, G. Salha-Galvan, R. Hennequin, T. Bouabça, T. Cazenave}

\begin{abstract}
A prevalent practice in recommender systems consists in averaging item embeddings to represent users or higher-level concepts in the same embedding space. This paper investigates the relevance of such a practice. For this purpose, we propose an expected precision score, designed to measure the consistency of an average embedding relative to the items used for its construction. We subsequently analyze the mathematical expression of this score in a theoretical setting with specific assumptions, as well as its empirical behavior on real-world data from music streaming services. Our results emphasize that real-world averages are less consistent for recommendation, which paves the way for future research to better align real-world embeddings with assumptions from our theoretical setting.
\end{abstract}


\maketitle

\section{Introduction}
\label{s1}

Modern recommender systems often leverage representation learning techniques to summarize similarities between recommendable items \cite{schedl2018current,okura2017embedding,yang2016survey,zhang2019deep}. These techniques learn low-dimensional vectorial representations of~these items, also known as \textit{embedding vectors} or simply \textit{embeddings}, in a common vector space where item proximity should reflect user preferences (for a \textit{collaborative filtering} system~\cite{koren2015advances}) or resemblance of item characteristics (for a \textit{content-based} system~\cite{javed2021review}). By computing {similarity metrics} such as the inner product or Euclidean distance between embeddings, the recommender system can subsequently identify new items similar to the ones each user has interacted with~\cite{al2018similarity,jain2020survey}.

A prevalent practice associated with the use of such embeddings in industry-oriented research and applications consists in \textit{averaging item embeddings} to obtain embeddings for users or higher-level concepts in the same vector space~\cite{briand2021semi,hansen2020contextual,pal2020pinnersage,wang2018billion,grbovic2018real,okura2017embedding}. 
As an illustration, Spotify learns embedding representations of listening sessions by averaging pre-computed embeddings of the music tracks listened to during these sessions~\cite{hansen2020contextual}.
This service also computes ``long-term'' user embeddings, used for recommendation purposes, by averaging the session embeddings associated with each user. Deezer computes embeddings for several types of recommendable music collections, such as playlists and albums, by averaging embeddings of the music tracks present in these collections~\cite{briand2021semi}. 
Yahoo averages embeddings of the news articles previously browsed by each user to represent them in the same embedding space as articles and provide personalized news recommendations~\cite{okura2017embedding}.
Alibaba averages side information embeddings, including category and brand embeddings, to obtain product embeddings for cold start recommendation on the Taobao e-commerce platform~\cite{wang2018billion}.

However, despite its prevalence, this averaging practice is often adopted without explicit justification from a theoretical standpoint.
As we detail in Section~\ref{s2}, the rationale for averaging item embeddings mainly stands in the simplicity and scalability of this approach~\cite{youtube,briand2021semi}. 
Yet, it is unclear to which extent averaging item embeddings guarantees to provide faithful user or higher-level concept representations for recommendation. 
For instance, assuming we represent a user by the average embedding of their previously consumed items, to what degree would the neighboring items of this average constitute relevant recommendations for this user? While the impact of averaging and other pooling operations has been studied in other fields, notably natural language processing (NLP)~\cite{blog,arora2017simple}, to our knowledge, the consistency of averaging operations remains relatively understudied in the specific context of a recommender~system.

In this short paper, we propose to investigate these important considerations, making the following contributions:
\begin{itemize}
    \item Firstly, we define $\text{Consistency}_k(\mathcal{X})$, a general expected precision score introduced in this study to measure the consistency of an average embedding relative to the items it summarizes, from a recommendation~standpoint.
        \item Secondly, we examine the consistency of averaging operations in a \textit{theoretical} setting with general assumptions on item embeddings, providing an in-depth analysis of the expression of $\text{Consistency}_k(\mathcal{X})$ in this setting.
        \item Thirdly, we analyze the \textit{empirical} behavior of this score on real-world data. Our experiments consider three variants of large-scale music track embeddings obtained from the music streaming service Deezer~\cite{briand2021semi}.
        \item Lastly, we discuss the discrepancies between our theoretical and empirical results, emphasizing that real-world averages are less consistent for recommendation. This discussion paves the way for future research to better align real-world data with our theoretical assumptions. Overall, we believe this study will be insightful for researchers and practitioners aiming to improve the faithfulness of average embeddings in their recommender systems.
\end{itemize}

This paper is organized as follows. In Section~\ref{s2}, we introduce the average embedding consistency problem more precisely. We subsequently present our proposed $\text{Consistency}_k(\mathcal{X})$ score. We report our theoretical analysis of this score in Section~\ref{s3}, and discuss our experimental results on real-world data in Section~\ref{s4}. Finally, we conclude in Section~\ref{s5}.

\section{Problem Formulation and Evaluation}
\label{s2}

\subsection{Preliminaries} 
\label{s21}
\subsubsection{Mathematical Notation} Throughout this paper, we consider  a set $\mathcal{X} = \{X_{i}\}_{1 \leq i \leq N}$ of $N \in \mathbb{N}^*$ real-valued vectors of dimension $d \in \mathbb{N}^*$, i.e., $\forall i \in \{1, \dots, N\}, X_i = (X_{i,1}, X_{i,2}, \dots, X_{i,d})^\intercal \in \mathbb{R}^{d}$.
We denote by $\mathcal{X}_k$ the set of subsets of $\mathcal{X}$ of cardinality $k$, for any $k \in \{1,\dots,N\}$. For any vector $z \in \mathbb{R}^{d}$ and $k \in \{1,\dots,N\}$, we define $\mathcal{X}_k(z) \in \mathcal{X}_k$ as the set of the $k$~\textit{nearest~neighbors}\footnote{This set might not be unique if $z$ is equidistant to several elements of $\mathcal{X}$, in which case $\mathcal{X}_k(z)$ can be drawn uniformly from all complying sets.} of $z$ among the $N$ elements of $\mathcal{X}$, according to some \textit{similarity metric} $s: \mathbb{R}^d \times \mathbb{R}^d \rightarrow \mathbb{R}$, i.e.,
\begin{equation}
   \mathcal{X}_k(z) =  \argmax_{\mathcal{Y} \in \mathcal{X}_k} \sum_{y \in \mathcal{Y}} s(z, y).
\end{equation}
Moreover, for any subset $\mathcal{U} = \{u_{i}\}_{1 \leq i \leq k} \subseteq \mathcal{X}$ of $k$ vectors, we define the \textit{center} or \textit{average} of $\mathcal{U}$ as $\mu_{\mathcal{U}} = \frac{1}{k}\sum_{u_{i} \in \mathcal{U}} u_{i}$,
and we denote the complementary set of $\mathcal{U}$ within $\mathcal{X}$ as $\overline{\mathcal{U}}$, i.e.,  $\overline{\mathcal{U}}= \mathcal{X}\setminus\mathcal{U}=\{X \in \mathcal{X}, X \notin\mathcal{U}\}$.

\subsubsection{From Mathematics to Recommender Systems} 
In the context of an embedding-based recommender system~\cite{zhang2019deep,koren2015advances}, $\mathcal{X}$ would correspond to a set of $d$-dimensional embedding representations associated with all recommendable \textit{items} in a catalog, while $\mathcal{U} \in \mathcal{X}_k$ would correspond to the embeddings of a subset of $k$ items. 
For instance, $\mathcal{X}$ could represent music tracks on a music streaming service, and $\mathcal{U}$ could represent the tracks present in a playlist or album of length~$k$~\cite{hansen2020contextual,briand2021semi}. Alternatively, $\mathcal{X}$ could represent all products from an e-commerce platform and $\mathcal{U}$ the shopping cart of a user~\cite{wang2018billion,wu2017session}. 
At this stage, we do not make assumptions regarding the representation learning technique used~to~learn~these~embeddings.

\subsection{Problem Formulation}
\label{s22}

\subsubsection{Averaging Embeddings} As illustrated in Section~\ref{s1}, recommender systems frequently average item embeddings.
Using the above notation, the action of averaging all embeddings of an item collection $\mathcal{U}$ to represent this collection translates to using $\mu_{\mathcal{U}}$ as the representation.
As $\mu_{\mathcal{U}} \in \mathbb{R}^d$, one can interpret this vector as a new embedding in the same vector space as items,
and, therefore, measure the similarity between $\mu_{\mathcal{U}}$ and other items using the similarity metric $s$.

\subsubsection{How Relevant is this Practice?}
\label{s222}
This paper aims to rigorously investigate the consistency of this averaging~practice. For instance, if we represent a user by the average embedding $\mu_{\mathcal{U}}$ of the items $\mathcal{U}$ this user has consumed or liked, do the items similar to $\mu_{\mathcal{U}}$ in the embedding space (according to $s$) also  constitute items that the user would like? In the same way, if $\mu_{\mathcal{U}}$ summarizes an album composed of the tracks in $\mathcal{U}$, are tracks similar to $\mu_{\mathcal{U}}$ also similar to tracks~in~$\mathcal{U}$? 

Intuitively, for $\mu_{\mathcal{U}}$ to be faithful to $\mathcal{U}$, we expect $\mu_{\mathcal{U}}$ to remain similar to the original items from $\mathcal{U}$. For instance, the user embedding should remain similar to the items the user has already liked, and the album embedding to the tracks contained in the album. For this reason, this paper focuses on the following specific research question: \textit{to which extent and under which conditions does $\mu_{\mathcal{U}}$ remain similar to items from $\mathcal{U}$, i.e., to the items used for its construction?}

\subsubsection{Related Work}
To our knowledge, this theoretical question remains understudied in recommendation. 
Existing research predominantly relied on averages for practical reasons.
Average embeddings are faster and simpler to compute than alternatives such as neural aggregations~\cite{he2017neural,hidasi2015session}. They have also been praised for their scalability, as they provide fixed-size representations independently of the number of items in $\mathcal{U}$~\cite{youtube,briand2021semi}. Yet, recent research pointed out some of their limitations, e.g., to represent heterogeneous or contextual preferences~\cite{hansen2020contextual,park2018collaborative,tran2021hierarchical}. 
We also acknowledge that the pros and cons of average embeddings have been studied for other applications, such as NLP tasks~\cite{mishra2019survey,arora2017simple,white2015well,blog}.  While being out of our scope (averaging for recommendation), these studies confirm the~importance~of~our~research~problem.


\subsection{Problem Evaluation}

To evaluate the consistency of $\mu_{\mathcal{U}}$ in accordance with our formulation in Section~\ref{s222}, we propose the following~score:
\begin{dmath}[compact]
\label{eq:consistency}
\text{Consistency}_k(\mathcal{X}) = \mathbb{E}_{\mathcal{U}\in\mathcal{X}_k} \Big[\text{Precision}_k(\mathcal{U})\Big], \allowbreak \text{ where } 
\text{Precision}_k(\mathcal{U})=\frac{|\mathcal{X}_k(\mu_{\mathcal{U}}) \bigcap \mathcal{U}|}{k} \in [0, 1],
\end{dmath}
for a given $k \in \{1,\dots,N\}$.
In essence, $\text{Precision}_k(\mathcal{U})$ measures the percentage of items from $\mathcal{U}$ among the $k$ nearest neighbors of $\mu_{\mathcal{U}}$ in $\mathcal{X}$. A perfect precision of 1 indicates that $\mathcal{X}_k(\mu_{\mathcal{U}}) = \mathcal{U}$.
Therefore, higher values of $\text{Consistency}_k(\mathcal{X})$ indicate that, on expectation, average embeddings computed from $\mathcal{X}$ will comprise more items used for their constructions in their neighborhood.
The remainder of this paper provides an analysis of $\text{Consistency}_k(\mathcal{X})$ in a theoretical setting with assumptions of the distribution of embeddings in $\mathcal{X}$, and studies its empirical behavior on real-world~embeddings.

\section{A Theoretical Analysis of $\text{CONSISTENCY}_k(\mathcal{X})$}
\label{s3}

We dedicate Section~\ref{s3} to our theoretical analysis of $\text{Consistency}_k(\mathcal{X})$. For clarity, we only present our setting, main results, and interpretation of these results in this section. We report all mathematical~proofs in Appendices~\ref{s32}~and~\ref{s33}.

\subsection{Setting and Assumptions} 
\label{s311} We focus on the setting where $\mathcal{X}$ is a set of independent and identically distributed (i.i.d.) multi-dimensional random variables (r.v.). For every $X_{i} \in \mathcal{X}$, the elements $\{X_{i,j}\}_{1 \leq j \leq d}$ form a set of i.i.d uni-dimensional r.v. and we denote by $\mu$, $\sigma^{2}$, $\gamma$, and $\kappa$ the mean, variance, skewness, and kurtosis of their distribution, respectively. We assume that these moments are finite. $s$ is the prevalent \textit{inner product similarity}: $\forall (x,y) \in \mathbb{R}^{d} \times \mathbb{R}^{d}, s(x, y) = x^\intercal y = \sum_{i=1}^{d}x_i y_i$.

\subsection{Main Results and Interpretations}
In this theoretical setting, we obtain the following results on average embeddings.

\begin{proposition}
\label{proposition1}
Let $k \in \{2,\dots,N\}$, $\mathcal{U} \in \mathcal{X}_k$, $u_{\text{in}} \in \mathcal{U}$, and $u_{\text{out}} \in \overline{\mathcal{U}}$. Then, under the hypotheses of Section~\ref{s311}, both $s_{\text{in}} = s(u_{\text{in}}, \mu_{\mathcal{U}})$ and $s_{\text{out}}=s(u_{\text{out}}, \mu_{\mathcal{U}})$ converge in probability to normal distributions as $d$ increases. Their respective means and variances are provided in the proof of Appendix~\ref{s32}.
Moreover, under such distributions, we have: 
\begin{dmath}[compact]
\label{eq:th_result_1}
    \mathbb{P}\Big(s(u_{\text{in}}, \mu_{\mathcal{U}}) > s(u_{\text{out}}, \mu_{\mathcal{U}})\Big)  =
    \frac{1}{2}\Big(1+\erf(\sqrt{\frac{d\sigma^{2}}{2((2(k-1)+\kappa)\sigma^{2} + 2k\gamma\mu\sigma+ 2k^{2}\mu^{2})}})\Big),
\end{dmath}
where $\erf$ denotes the Gauss error function~\cite{andrews1998special}: $\erf : x \mapsto \frac{2}{\sqrt{\pi}}\int_{0}^{x}e^{-t^2}dt$.
\end{proposition}

\begin{figure*}[t!]
\begin{center}
    
\begin{subfigure}{.325\linewidth}
\includegraphics[width=\linewidth]{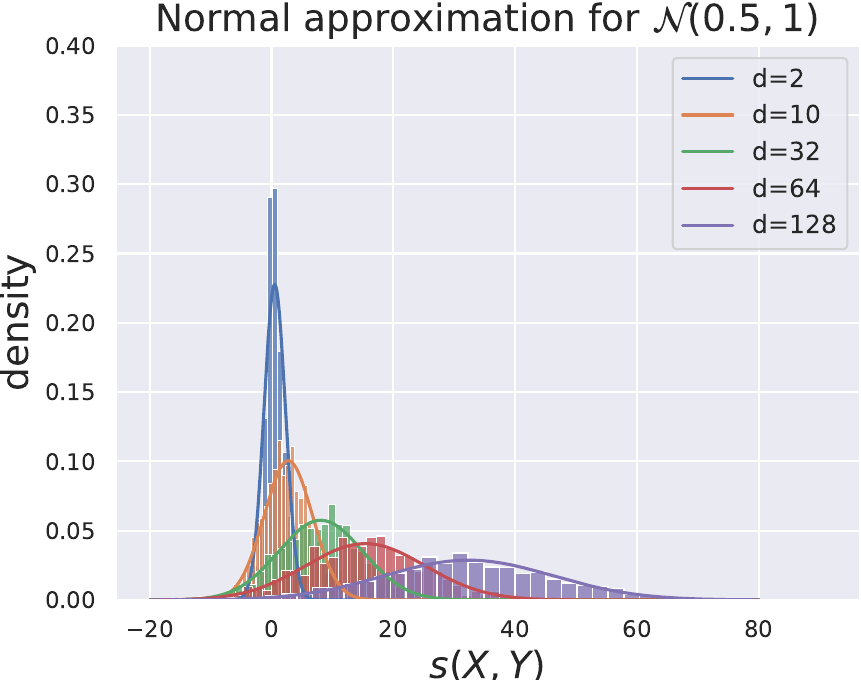}
  \caption{}
  \label{fig1a}
\end{subfigure} \hfill 
\begin{subfigure}{.32\linewidth}
  \includegraphics[width=\linewidth]{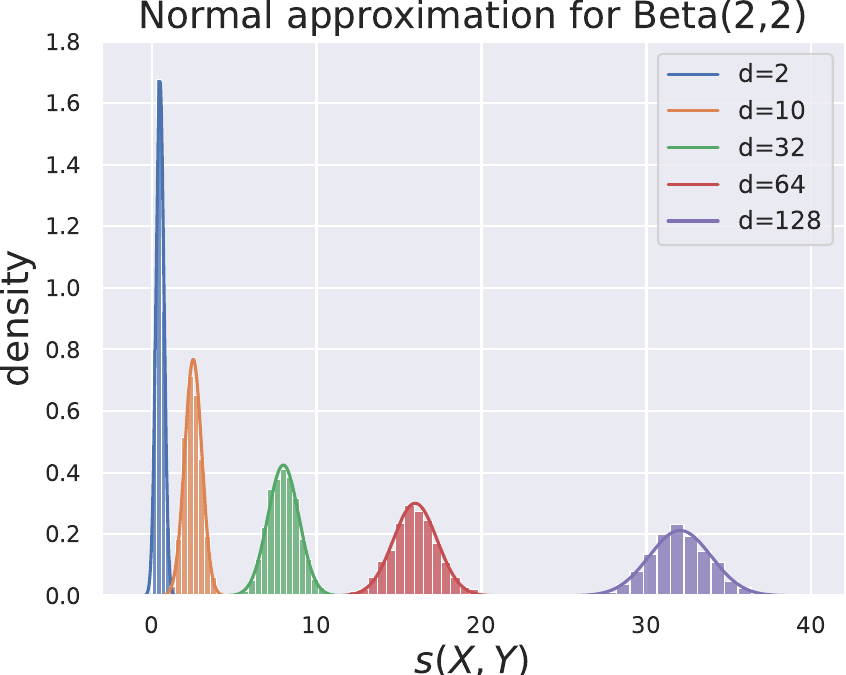}
  \caption{}
  \label{fig1b}
\end{subfigure}\hfill 
\begin{subfigure}{.32 \linewidth}
  \includegraphics[width=\linewidth]{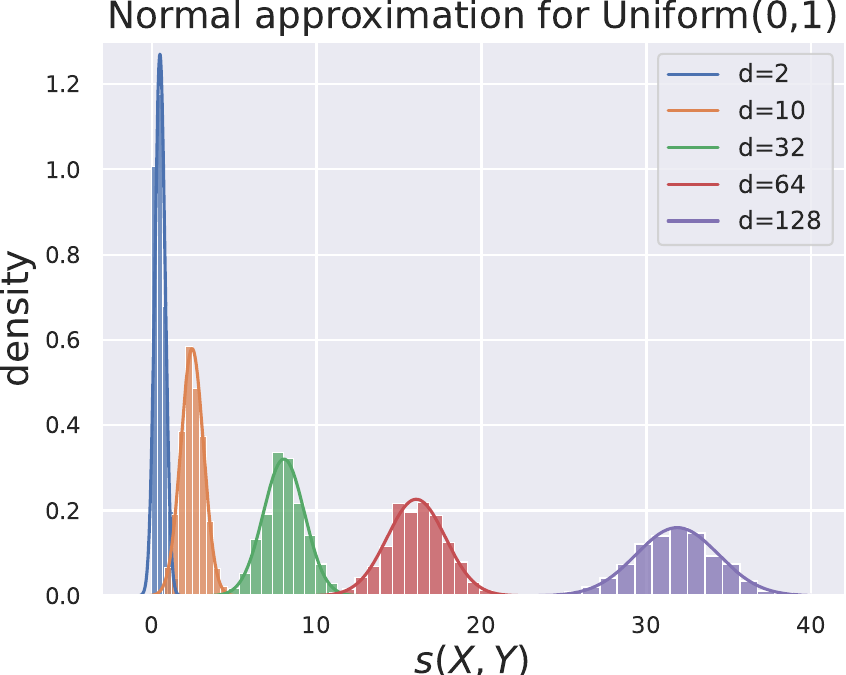}
  \caption{}
  \label{fig1c}
\end{subfigure}

\caption{Histograms of inner product similarity values between 1~000 $d$-dimensional vectors with entries randomly drawn from $\mathcal{N}(0.5,1)$ (Figure~\ref{fig1a}), $\text{Beta}(2,2)$ (Figure~\ref{fig1b}), or $\text{Uniform}(0,1)$ (Figure~\ref{fig1c}) distributions, with $d \in \{2,10,32,64,128\}$. Histograms undergo a rightward shift as $d$ increases, since similarity computations involve summing more elements (see Section~\ref{s311}). Curves correspond to normal approximations of similarity distributions. We observe that, for the largest values of $d$, the normal distribution faithfully approximates all similarity distributions, an important result to validate the approximations of Propositions~\ref{proposition1}~and~\ref{proposition2}.}

\label{figure1}
\end{center}
\end{figure*}

Proposition~\ref{proposition1} is a consequence of the \textit{central limit theorem}~\cite{fischer2011history} applied to inner product similarities. 
Figure~\ref{figure1} illustrates that, for dimension values such as $d=128$ (a common choice in recommendation applications), the normal distribution faithfully approximates the distribution of similarities, regardless of the original embedding distribution. 
Hence, Equation~\eqref{eq:th_result_1} provides a reliable approximation of the probability that an element from $\mathcal{U}$ will be closer to its center $\mu_{\mathcal{U}}$ than a random point from $\overline{\mathcal{U}}$.
It is worth noting that the probability is guaranteed to be greater than 0.5, as the expression within the erf function is positive.
We also observe that the probability increases with $d$,
illustrating that, as the dimension increases, the chances for a vector from $\overline{\mathcal{U}}$ to have a stronger similarity with $\mu_{\mathcal{U}}$ than elements~of~$\mathcal{U}$~diminishes. 


The case of centered embeddings ($\mu=0$) is of particular interest. Indeed, for $\mu=0$ and a fixed $d$, the probability becomes independent of $\sigma$ and $\gamma$, and a decreasing function of $k$ and $\kappa$. 
In essence, as $k$ increases, it becomes increasingly difficult for $\mu_{\mathcal{U}}$ to remain similar to all $\mathcal{U}$ items, while remaining dissimilar to all $\overline{\mathcal{U}}$ items. The decrease with respect to the kurtosis $\kappa$ relates to its interpretation as a measure of the propensity of a distribution to produce outliers~\cite{westfall2014kurtosis}. Intuitively, the presence of outliers would impact the ability for $\mu_{\mathcal{U}}$ to remain similar to all $\mathcal{U}$ items and dissimilar to all $\overline{\mathcal{U}}$ items\footnote{On the contrary, Equation~\eqref{eq:th_result_1} would be maximized by distributions with minimal kurtosis, such as a re-centered $\text{Bernoulli}(0.5)$ or a Rademacher distribution ($X_{i,j} = 1$ or $-1$ with probability $0.5$ each). We emphasize that these distributions are \textit{discrete}, and might therefore allow for less precise similarity computations between vectors. This brings to light an interesting \textit{trade-off} between similarity precision and average embedding consistency.}. In the following, we continue to focus on $\mu = 0$, using results from Proposition~\ref{proposition1} to express~$\text{Consistency}_k(\mathcal{X})$.
 



\begin{proposition}
\label{proposition2}
 Under the hypotheses of Section~\ref{s311} and approximated distributions of Proposition~\ref{proposition1} with $\mu=0$: 
\begin{equation}
\label{eq:th_result_2}
    \text{Consistency}_k(\mathcal{X})  
 = \frac{1}{k} \sum_{i=1}^{k}\int_{-\infty}^{\infty}f_{\text{in},(i)}(x)\times F_{\text{out},(k-i+1)}(x)\d x,
\end{equation}
    where explicit formulas for the $f_{\text{in},(i)}$ and $F_{\text{out},(k-i+1)}$ functions are provided in the proof of Appendix~\ref{s33}.
\end{proposition}



\begin{figure*}[t!]
\begin{center}
\begin{subfigure}{.32\linewidth}
\includegraphics[width=\linewidth]{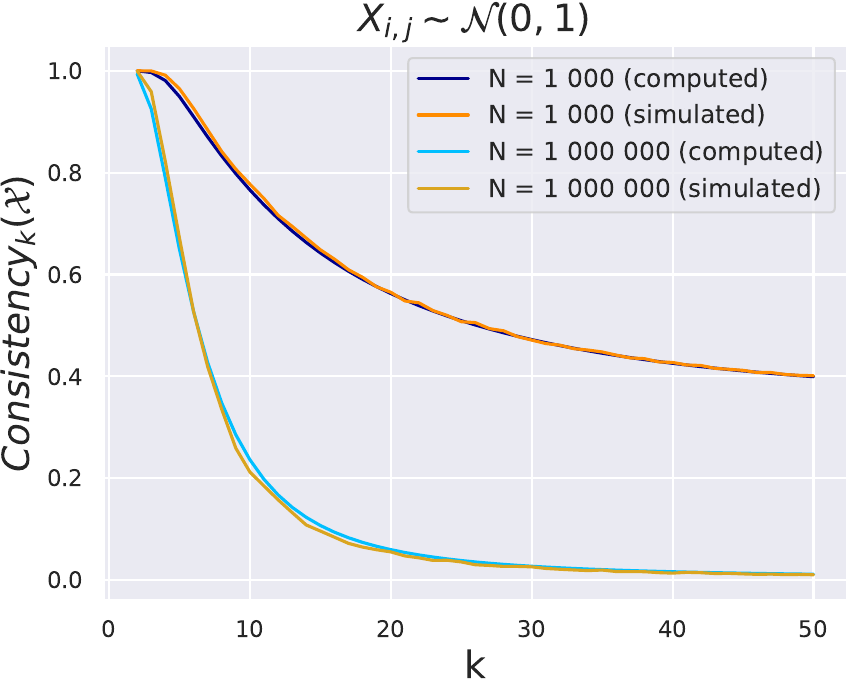}
  \caption{}
  \label{fig2a}
\end{subfigure} \hfill 
\begin{subfigure}{.32\linewidth}
  \includegraphics[width=\linewidth]{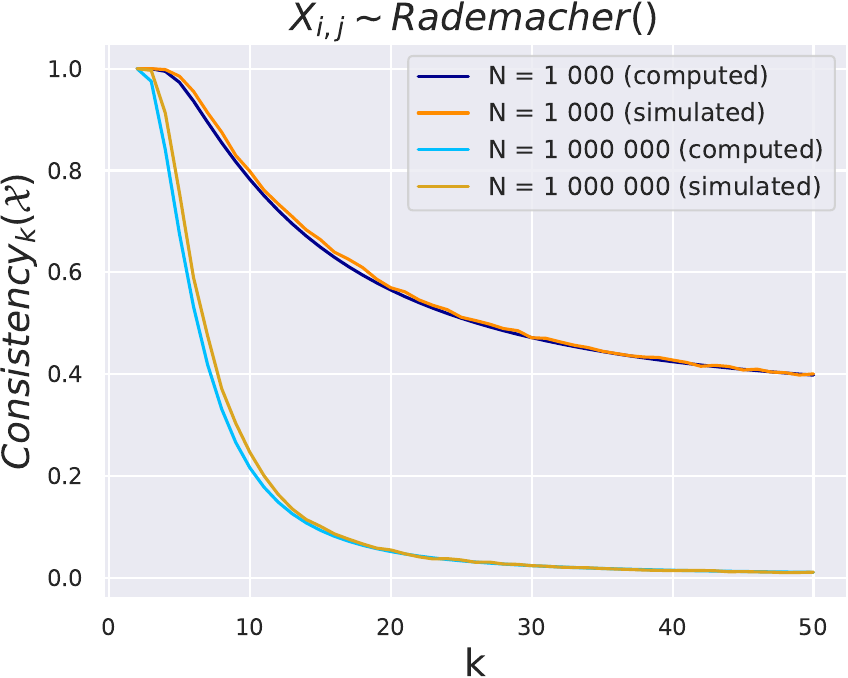}
  \caption{}
  \label{fig2b}
\end{subfigure}\hfill 
\begin{subfigure}{.32 \linewidth}
  \includegraphics[width=\linewidth]{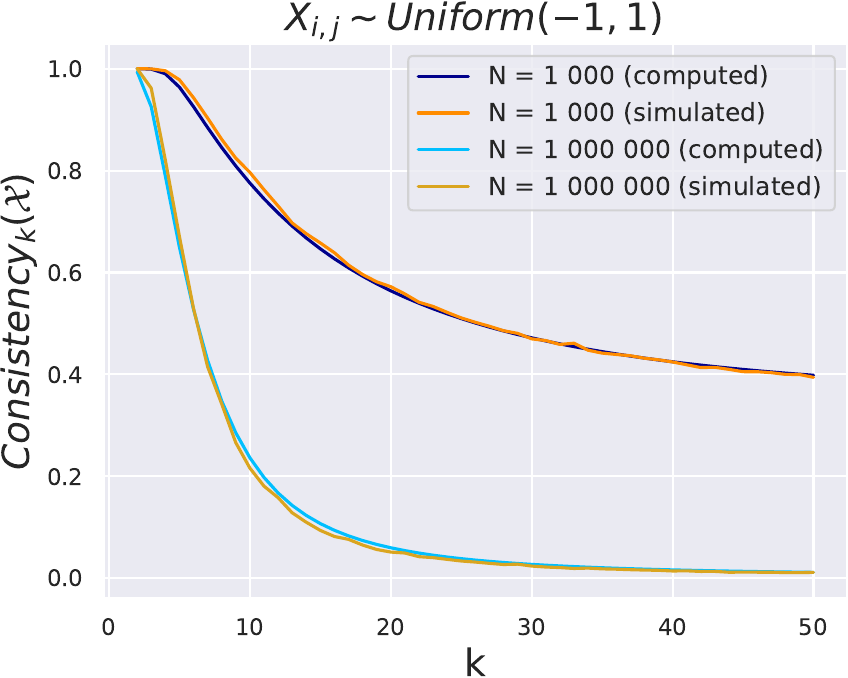}
  \caption{}
  \label{fig2c}
\end{subfigure}

\caption{Comparison of the $\text{Consistency}_k(\mathcal{X})$ scores obtained by a direct \textit{computation} of Equation~\eqref{eq:th_result_2} to the ones estimated via numerical \textit{simulations}, for $k \in \{2, \dots, 50\}$, with $d =$ 128, $N =$ 1~000 or 1~000~000, and $X_{i,j} \sim \mathcal{N}(0,1)$ (Figure~\ref{fig2a}), $X_{i,j} \sim \text{Rademacher}()$ (Figure~\ref{fig2b}), or $X_{i,j} \sim \text{Uniform}(-1,1)$ (Figure~\ref{fig2c}).
We used the Python library \textit{scipy} to compute integrals from Equation~\eqref{eq:th_result_2}. For numerical simulations, we sampled $N$ vectors from the above three distributions. Then, for each $k$, we randomly picked a subset of $k$ vectors and computed the precision of this subset (Equation~\ref{eq:consistency}). We repeated this operation 1~000 times for each value of $k$ and reported averaged scores on figures. 
}
\label{figure2}
\end{center}
\end{figure*}

Proposition~\ref{proposition2} provides a useful approximated analytical expression of $\text{Consistency}_k(\mathcal{X})$. In
Figure \ref{figure2}, we assess the accuracy of this expression by comparing, for various $X_{i,j}$ distributions and values of $k$, the $\text{Consistency}_k(\mathcal{X})$ score obtained from Equation~\eqref{eq:th_result_2} to the one estimated via numerical simulations.
Our computed expression systematically coincides with the simulated score, validating the correctness of Proposition~\ref{proposition2} and the relevance of our approximations. 

Overall, we observe that $\text{Consistency}_k(\mathcal{X})$ decreases with $k$ and the number of items $N = |\mathcal{X}|$. Regarding $k$, this result is coherent with our above interpretation of Proposition~\ref{proposition1}. Regarding $N$, such a result is also unsurprising. Indeed, the more vectors in $\mathcal{X}$, the greater the likelihood of some unrelated item embeddings becoming similar to $\mu_{\mathcal{U}}$ by~chance.

Yet, in Figure~\ref{figure2}, all scores  
remain quite high for $N=$~1~000 (e.g., around 0.4 for $k=50$). Importantly, for small values of $k$, the consistency of average embeddings remains close to 1 even for $N=$~1~000~000. 
Therefore, even with a large catalog of millions of items, as in music streaming services, averaging item embeddings appears as a consistent way to faithfully represent collections of a few items, provided that these embeddings comply with our~theoretical~assumptions.

 \section{From Theory to Practice}
\label{s4}
\subsection{Experimental Setting}
In this Section~\ref{s4}, we analyze the empirical behavior of $\text{Consistency}_k(\mathcal{X})$ on real-world data, with the aim of discussing potential discrepancies with our results from Section~\ref{s3}. Our experiments focus on three \textit{music track embeddings} datasets. As illustrated in this paper, music recommendation is especially prone to high-order concept learning: album, playlist, session, music genre, and user embeddings can all be obtained by averaging music track embeddings~\cite{briand2021semi,hansen2020contextual,schedl2018current,bendada2023scalable}.

Firstly, we consider two variants\footnote{We release our source code on GitHub to ensure the reproducibility of our experimental analysis on these two Deezer public datasets: \href{https://github.com/deezer/consistency}{https://github.com/deezer/consistency}.} of 50~000 music track embeddings publicly released by Deezer~\cite{briand2021semi}.
The first ones, denoted \textit{TT-SVD embeddings}, consist of 128-dimensional vectors obtained by factorizing a track-track (TT) pointwise mutual information matrix computing track co-occurrences in Deezer playlists, using singular value decomposition (SVD)~\cite{briand2021semi}. The second ones, denoted \textit{UT-ALS embeddings}, are 256-dimensional vectors obtained by factorizing a user-track (UT) interaction matrix, using alternating least squares~(ALS)~\cite{briand2021semi}.

Besides, we report results on \textit{2M-TT-SVD embeddings}, a private dataset of two million 128-dimensional track embeddings, extracted from Deezer's production environment. These embeddings are computed using SVD on a co-occurrence matrix comparable to the TT-SVD one. We voluntarily omit technical details for confidentiality~reasons. 

\label{s41}
\subsection{Results and Discussion}
\label{s42}

\begin{figure*}[t!]
\begin{center}
    
\begin{subfigure}{.32\linewidth}
  \includegraphics[width=\linewidth]{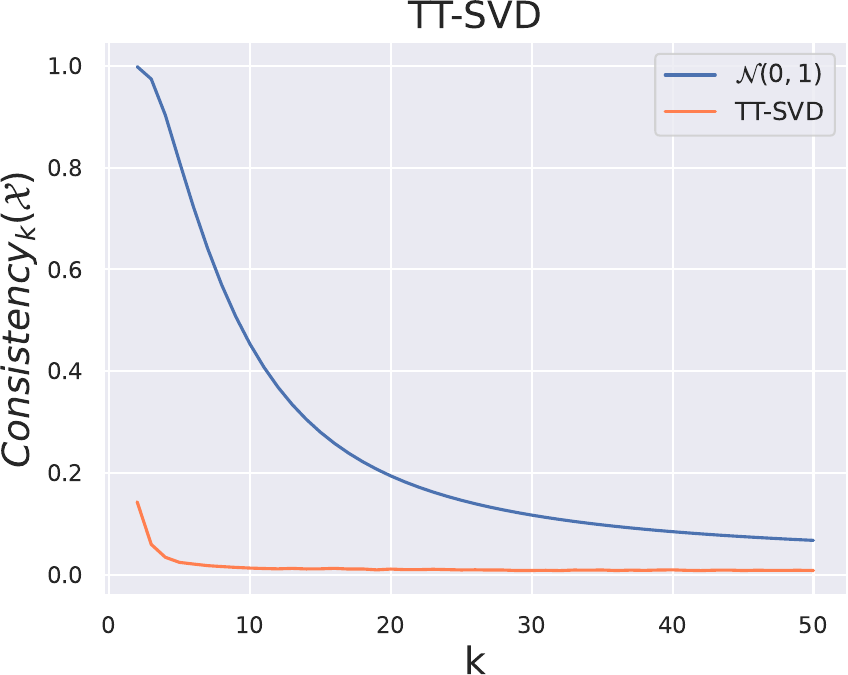}
  \caption{}
  \label{fig3a}
\end{subfigure}  \hfill 
\begin{subfigure}{.32\linewidth}
  \includegraphics[width=\linewidth]{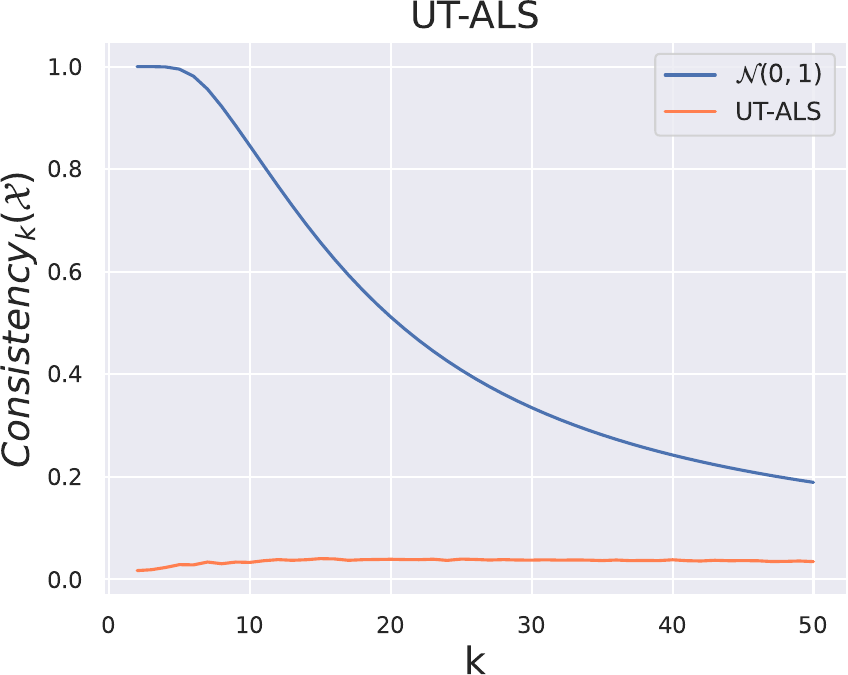}
  \caption{}
  \label{fig3b}
\end{subfigure}  \hfill
\begin{subfigure}{.32\linewidth}
  \includegraphics[width=\linewidth]{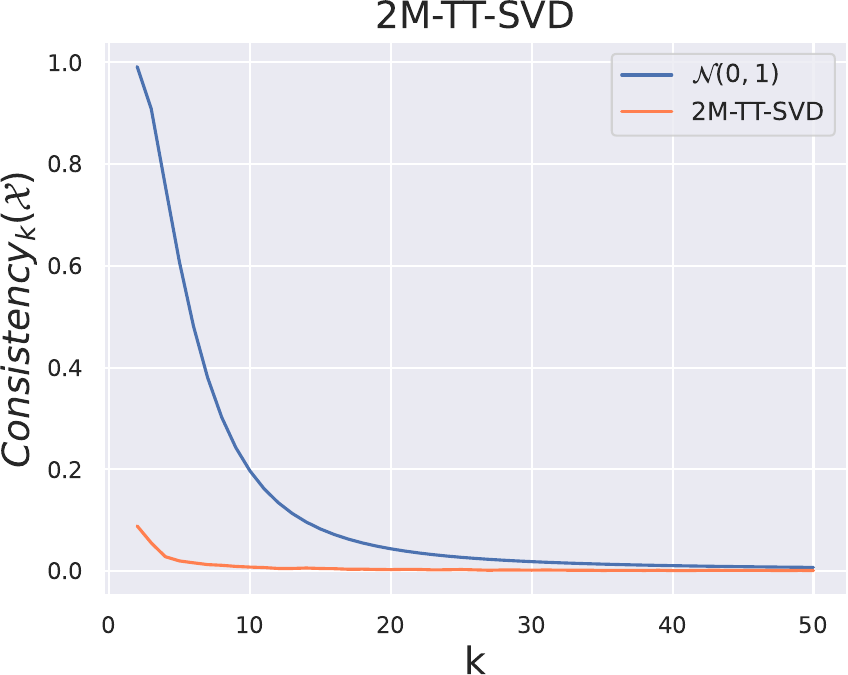}
  \caption{}
  \label{fig3c}
\end{subfigure}

\caption{$\text{Consistency}_k(\mathcal{X})$ scores of centered versions of TT-SVD (Figure~\ref{fig3a}), UT-ALS (Figure~\ref{fig3b}), and 2M-TT-SVD embeddings (\ref{fig3c}), for $k\in\{2,\dots,50\}$ and the inner product similarity $s$. For comparison, figures also display the $\text{Consistency}_k(\mathcal{X})$ scores of embeddings generated from normal distributions, with the same dimension and number of items as the real-world embeddings under consideration. Average embeddings of real-world data are less consistent than those of data complying with our theoretical setting~from~Section~\ref{s3}.}
\label{figure3}
\end{center}
\end{figure*}

Figure~\ref{figure3} reports our evaluation of $\text{Consistency}_k(\mathcal{X})$ scores for the three music track embeddings under consideration. 
Our experiments show that average embeddings of real-world data are less consistent for recommendation than those computed from embedding data explicitly complying with our theoretical setting from Section~\ref{s3}.

Specifically, one can not ensure that the average embedding of a collection of items will remain similar to the items present in this collection. Even for low values of $k$, the consistency of average embeddings does not surpass 14\%,  6\%, and 2 \% for TT-SVD, 2M-TT-SVD, and UT-ALS embeddings, respectively. 
Consequently, even the average of two randomly selected embeddings would likely result in a vector whose two most similar neighbors will not be the selected embeddings~themselves.

Regarding SVD-based embeddings (TT-SVD, 2M-TT-SVD), we observe that consistency scores drop as $k$ increases, as in Figure~\ref{figure2}. On the contrary, scores remain steady at around 2\% for UT-ALS embeddings. 
This phenomenon highlights the dissimilarity in distributions of embeddings generated by different representation learning algorithms (SVD, ALS). In particular, the steady consistency of UT-ALS suggests that these embeddings might be more suitable for downstream applications involving average operations on large collections, while TT-SVD embeddings might be preferable for applications with low values of $k$,
although more studies would be required in future work~for~confirmation.

Overall, our experiments also pave the way for future work to align real-world embeddings with the setting from Section~\ref{s3}, to improve the consistency of real-world averages.
For instance, one could consider alternating SVD or ALS matrix reconstruction optimization steps with projections within the set of distributions complying with assumptions from Section~\ref{s3}.
One could also examine adding a regularization term to the optimized loss during training, e.g., the Kullback-Leibler divergence of embeddings with a pre-selected complying distribution. In particular, these strategies could help to enforce the identical distribution of embedding dimensions.
In addition, we note that, in Figure~\ref{figure3}, we computed scores on \textit{centered} embeddings. Besides being in line with Section~\ref{s3}, our tests revealed that this centering operation slightly improves the consistency of TT-SVD, UT-ALS, and 2M-TT-SVD average embeddings (albeit modifying initial similarities). Our future research will aim to further understand the impact of centering embeddings on consistencies.


\section{Conclusion}
\label{s5}

This short paper proposed a rigorous study of the common practice consisting of averaging item embeddings in recommender systems. We provided a mathematical analysis of the consistency of these averaging operations in a general theoretical setting, as well as an empirical evaluation on real-world data. Our results revealed that real-world averages were less consistent than those computed in our theoretical setting. This sets the stage for future research directions, discussed in this paper, toward better aligning real-world data with our theoretical assumptions. Due to the prevalence of the embedding averaging practice in industry-oriented research and applications, we believe our study and proposed directions will be insightful for the recommendation community, and could eventually lead to the improvement of embedding-based recommender systems leveraging average embeddings for representation learning.

\clearpage
\appendix

\section*{Appendix}

In this supplementary section, we report the mathematical proofs of our Propositions~\ref{proposition1}~and~\ref{proposition2} from Section~\ref{s3}.
\section{Proof of Proposition~\ref{proposition1}}
\label{s32}
\subsection{Preliminaries}
Let $X$, $Y$, and $Z$ be $d$-dimensional random variables (r.v.) composed of $d$ independent and identically distributed (i.i.d.) uni-dimensional r.v. as elements. Distributions might differ between $X$,$Y$, and $Z$. Let $\mu_{X}, \sigma^{2}_{X}, \gamma_{X}, \kappa_{X}$ be the finite mean, variance, skewness, and kurtosis of the distribution of elements of $X$, respectively. Let $\mu_{Y}, \sigma^{2}_{Y}, \mu_{Z}, \sigma^{2}_{Z}$ be the finite mean and variance of the elements of $Y$ and $Z$, respectively. When $d$ increases, the following approximations hold:
\begin{equation}
\begin{split}
s(X,Y) &\sim \mathcal{N}(\mu_{XY}, \sigma_{XY}^{2}), \\
 s(X,X) &\sim \mathcal{N}(\mu_{XX}, \sigma_{XX}^{2}), 
\label{gaussian}
\end{split}
\end{equation}
with:
\begin{align*}
\mu_{XY} &= d \times \mu_{X}\mu_{Y}, \\
\sigma_{XY}^{2} &= d \times ((\sigma_{X}^{2} + \mu_{X}^{2}) (\sigma_{Y}^{2} + \mu_{Y}^{2}) -\mu_{X}^{2}\mu_{Y}^{2}), \\
\mu_{XX} &= d \times (\mu^{2}_{X} + \sigma^{2}_{X}), \\
\sigma_{XX}^{2} &= d \times (4\mu_{X}^{2}\sigma_{X}^{2}+ 4\mu_{X}\gamma_{X}\sigma_{X}^{3}+ (\kappa_{X} - 1)\sigma_{X}^4).
\end{align*}
Indeed, we have $s(X,Y) = \sum_{i=1}^dX_{i}Y_{i}$. Each r.v. $X_{i}Y_{i}$ verifies $\mathbb{E}[X_{i}Y_{i}] = \mu_{X}\mu_{Y}$ and $\Var(X_{i}Y_{i})=(\sigma_{X}^{2}+\mu_{X}^{2})(\sigma_{Y}^{2}+\mu_{Y}^{2}) - \mu_{X}^{2}\mu_{Y}^{2}.$
$s(X,Y)$ being the sum of $d$ i.i.d. r.v., according to the central limit theorem~\cite{fischer2011history}, it can be approximated by a normal distribution of expectation $d \times \E[X_{i}Y_{i}]$ and variance $d \times \Var(X_{i}Y_{i})$ for large values of $d$, leading to the approximation for $s(X,Y)$ in Equation~\eqref{gaussian}. A similar reasoning leads to the approximation~for~$s(X,X)$ in this same Equation~\eqref{gaussian}. \newline

\noindent Moreover, regarding covariances of these similarities, we have: 
\begin{equation}
\begin{split}
\Cov(s(X,Y), s(X,Z)) &= d\times\sigma_{X}^{2}\mu_{Y}\mu_{Z} ,\\ \Cov(s(X,X), s(X,Y)) &= d\times\mu_{Y}(\gamma_{X}\sigma_{X}^{3}+ 2\mu_{X}\sigma_{X}^{2}).
\label{covariance}
\end{split}
\end{equation}
Indeed: \begin{align*}
    \Cov(s(X,Y), s(X,Z)) &= \mathbb{E}[s(X,Y)s(X,Z)] -\mathbb{E}[s(X,Y)]\mathbb{E}[s(X,Z)] \\
    &=\mathbb{E}[(\sum_{i=1}^{d}X_{i}Y_{i})(\sum_{j=1}^{d}X_{j}Y_{j})] - d \times \mu_{X}\mu_{Y} \times d \times \mu_{X}\mu_{Z} \\
    &= \allowbreak \mathbb{E}[\sum_{i=1}^{d}(X_{i}Y_{i}X_{i}Z_{i} + \sum_{j=1 \atop j\neq i}^{d}X_{i}Y_{i}X_{j}Z_{j})] - d^2 \times \mu_{X}^2\mu_{Y}\mu_{Z} \\
    &= \allowbreak \sum_{i=1}^{d} (\mathbb{E}[X_{i}^2]\mathbb{E}[Y_{i}]\mathbb{E}[Z_{i}] \allowbreak + \sum_{j=1 \atop j\neq i}^{d}(\mathbb{E}[X_{i}]  \mathbb{E}[Y_{i}] \mathbb{E}[X_{j}] \mathbb{E}[Z_{j}]))   - d^2 \times \mu_{X}^2\mu_{Y}\mu_{Z} \\ &= d \times((\sigma_{X}^2+\mu_{X}^2)\mu_{Y}\mu_{Z} + (d-1)(\mu_{X}^2\mu_{Y}\mu_{Z}))- d^2 \times \mu_{X}^2\mu_{Y}\mu_{Z} \\
    &= d \times \sigma_{X}^2\mu_{Y}\mu_{Z}.\end{align*}
We obtain the second covariance of Equation~\eqref{covariance} with similar computations. Notice how $\mu_{Y}=0$ or $\mu_{Z}=0$ implies that similarities are uncorrelated and, therefore, also independent when they follow multivariate normal distributions~\cite{jacod2004probability}.

\subsection{Distribution of $s_{\text{in}} = s(u_{\text{in}}, \mu_{\mathcal{U}})$ and $s_{\text{out}}=s(u_{\text{out}}, \mu_{\mathcal{U}})$} Using the distributive property of the inner product $s$, we~have:
    $$s_{\text{out}} = \frac{\sum\limits_{i=1}^{k}s(u_{\text{out}}, u_{i})}{k},$$ 
    i.e., the sum of $k$ correlated identically distributed normal distributions. The expectation of the sum is thus: 
    $$\mathbb{E}[s_{\text{out}}]=d\times\mu^{2}.$$
    We compute its variance using Bienaymé's identity~\cite{loeve1977probability} stating that, for $k$ r.v. $(A_{i})_{1 \leq i \leq k}$, then:
$$\Var(\sum_{i=1}^k A_i)=\sum_{i=1}^k \Var(A_i)+\sum_{i,j=1 \atop i \neq j}^k \Cov(A_i,A_j)=\sum_{i,j=1}^k\Cov(A_i,A_j).$$ Also considering $A_{i}=s(u_{\text{out}},u_{i})$, and using the first half of both Equations~\eqref{gaussian}~and~\eqref{covariance}, we obtain:
\begin{align*}
\Var(s_{\text{out}}) &= \frac{kd(\sigma^{4}+ 2\sigma^{2}\mu^{2})+ k(k-1)d\sigma^{2}\mu^{2}}{k^2} \\
&= d\frac{(\sigma^{4}+ (k+1)\sigma^{2}\mu^{2})}{k}.\end{align*} \newline

\noindent Besides, under preliminary approximations: 
$$s_{\text{in}} = \frac{s(u_{\text{in}}, u_{\text{in}})+\sum_{i\neq \text{in}} s(u_{\text{in}}, u_{i})}{k}$$ is the sum of $k$ correlated normal distributions, i.e., a normal distribution with:
$$\mathbb{E}[s_{\text{in}}]=d(\mu^{2}+\frac{\sigma^{2}}{k}).$$
Using Bienaymé's identity and Equations~\eqref{gaussian}~and~\eqref{covariance}, we get:

\begin{align*}
\Var(s_{\text{in}}) 
&= d \times \frac{4\mu^{2}\sigma^{2}+4\mu\gamma\sigma^{3}+(\kappa-1)\sigma^{4}+ (k-1)(\sigma^{4}+2\sigma^{2}\mu^{2})+2(k-1)\mu(\gamma\sigma^{3}+2\mu\sigma^{2})+(k-1)(k-2)\sigma^{2}\mu^2}{k^2} \\
&=\frac{d\times(k(k+3)\mu^{2}\sigma^{2}+4\mu\gamma\sigma^{3}+ (\kappa+k-2)\sigma^{4})}{k^2}.   
\end{align*}

\subsection{Distribution of
$s_{\text{diff}} = s_{\text{in}} - s_{\text{out}}$}
Using  preliminary approximations, and since the difference of two normally distributed r.v. is also normally distributed~\cite{jacod2004probability}, we obtain that $s_{\text{diff}}= s(u_{\text{in}}, \mu_{\mathcal{U}}) - s(u_{\text{out}}, \mu_{\mathcal{U}})$ follows a normal distribution with:
\begin{align*}
\mathbb{E}[s_{\text{diff}}] &=\mathbb{E}[s(u_{\text{in}}, \mu_{\mathcal{U}})] -\mathbb{E}[s(u_{\text{out}}, \mu_{\mathcal{U}})] \\ &= d\frac{\sigma^2}{k}.\end{align*}
To compute its variance, we remark that:
\begin{align*}
s_{\text{diff}} &= s(u_{\text{in}}, \mu_{\mathcal{U}}) -s(u_{\text{out}}, \mu_{\mathcal{U}}) \\ 
&= \frac{1}{k}(s(u_{\text{in}},u_{\text{in}}) + s(u_{\text{in}},\sum_{j\neq \text{in}}^{k}u_{j}) -s(u_{\text{out}},u_{\text{in}}) -s(u_{\text{out}},\sum_{j\neq \text{in}}u_{j})) \\
&= \frac{s(X,X)+s(X,Z)+s(Y,X)+s(Y,Z)}{k}, \end{align*} with $X=u_{\text{in}}$, $Y=-u_{\text{out}}$ and $Z=\sum_{j\neq \text{in}}u_{j}.$
Since $X$, $Y$, and $Z$ are three independent multidimensional normal distributions, for which dimensions are i.i.d. with expectation $\mu, -\mu, (d-1)\mu$ and variance $\sigma^{2}, \sigma^{2}, (d-1)^2\sigma^{2}$, we once again use Bienaymé's identity with Equations~\eqref{gaussian}~and~\eqref{covariance} to obtain:
     $$\Var(s_{\text{diff}})=d\frac{(2(k-1)+\kappa)\sigma^{4} + 2k\gamma\mu\sigma^{3}+ 2k^{2}\sigma^{2}\mu^{2}}{k^2}.$$ 
Finally, using the cumulative distribution function of normal distributions~\cite{jacod2004probability}, we get Equation~\eqref{eq:th_result_1}: 

\begin{align*}
&\mathbb{P}(s_{\text{diff}} > 0) = 1-\mathbb{P}(s_{\text{diff}}\leq0) = \frac{1}{2}(1-\erf(-\frac{\mathbb{E}[s_{\text{diff}}]} {\sqrt{2\Var(s_{\text{diff}})}})) \\ \iff
&\mathbb{P}(s(u_{\text{in}}, \mu_{\mathcal{U}}) > s(u_{\text{out}}, \mu_{\mathcal{U}})) = \frac{1}{2}(1+\erf(\sqrt{\frac{d\sigma^{2}}{2((2(k-1)+\kappa)\sigma^{2} + 2k\gamma\mu\sigma+ 2k^{2}\mu^{2})}})).
\end{align*} \newline


\section{Proof of Proposition~\ref{proposition2}}
\label{s33}
\begin{figure}[t]
  \centering
  \includegraphics[width=0.4\textwidth]{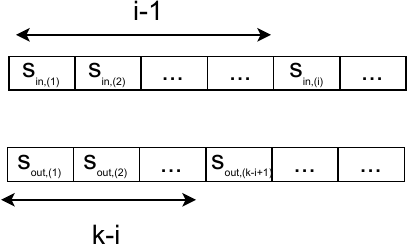}
  \caption{Ranking $s_{\text{in}}$ and $s_{\text{out}}$ statistics.}
  \label{fig:appendix}
\end{figure}

This proof starts by evaluating $p_{(\frac{i}{k})} = \mathbb{P}(\text{Precision}_k(\mathcal{U})) = \frac{i}{k})$ for $i \in \{1,\dots,k\}$, and then computes $\text{Consistency}_k(\mathcal{X}) = \mathbb{E}[\text{Precision}_k(\mathcal{U})] = \sum_{i=1}^{k}\frac{i}{k}\times p_{(\frac{i}{k})}$ under the studied hypotheses. To begin with, we set:
$$p_{(\frac{i}{k})}^{+} = \mathbb{P}(\text{Precision}_k(\mathcal{U}) \geq \frac{i}{k}), \text{ for } i \in \{1,\dots,k+1\}.$$
As $i$ takes integer values:
$$p_{(\frac{i}{k})} = p_{(\frac{i}{k})}^{+} -p_{(\frac{i+1}{k})}^{+}.$$ 
Assuming $p_{(\frac{k+1}{k})}^{+} = 0$ by definition, we have:
\begin{equation}
\mathbb{E}\Big[\text{Precision}_k(\mathcal{U})\Big] = \sum_{i=1}^{k}\frac{i}{k}\times (p_{(\frac{i}{k})}^{+} -p_{(\frac{i+1}{k})}^{+}) = \frac{1}{k}\sum_{i=1}^{k}p_{(\frac{i}{k})}^{+}.
\label{eq-proof2-1}
\end{equation}
Let us denote by $s_{\text{in},(i)}$ the distribution of the $i$\up{th} highest value of $S_{\text{in}}=\{s(u_{\text{in}}, \mu_{\mathcal{U}}), u_{\text{in}}\in\mathcal{U}\}$, and by $s_{\text{out},(k-i+1)}$ the $(k-i+1)$\up{th} highest value of $S_{\text{out}}=\{s(u_{\text{out}}, \mu_{\mathcal{U}}), u_{\text{out}}\in\overline{\mathcal{U}}\}$, for $i \in \{1,\dots,k\}$. Using this notation, we observe that:
$$p_{(\frac{i}{k})}^{+} = \mathbb{P}(s_{\text{in},(i)} > s_{\text{out},(k-i+1)}).$$  Indeed, as illustrated in Figure~\ref{fig:appendix}, we can show that $s_{\text{in},(i)} > s_{\text{out},(k-i+1)} \iff \text{Precision}_k(\mathcal{U})~\geq~\frac{i}{k}$:
\begin{itemize}
    \item If $s_{\text{in},(i)} > s_{\text{out},(k-i+1)}$ then at most $k-i$ elements of $S_{\text{out}}$ are greater than $s_{\text{in},(i)}$. Also, by definition, exactly $i-1$ elements of $S_{\text{in}}$ are greater than $s_{\text{in},(i)}$. So, at most $k-1$ similarities are greater than $s_{\text{in},(i)}$ overall. Thus, $s_{\text{in},(i)}$ is one of the $k$ highest values of $S_{\text{in}} \bigcup S_{\text{out}}$ and $\text{Precision}_k(\mathcal{U}) \geq \frac{i}{k}$.
    \item If $\text{Precision}_k(\mathcal{U}) \geq \frac{i}{k}$, then at least $i$ elements of $S_{\text{in}}$ are in the top-$k$ of $S_{\text{in}}\bigcup S_{\text{out}}$, including $s_{\text{in},(i)}$. That leaves at most $k-i$ slots available in the top-$k$. Since, by definition, exactly $k-i$ elements of $S_{\text{out}}$ are greater than $s_{\text{out},(k-i+1)}$, $s_{\text{out},(k-i+1)}$ is necessarily outside of the top-$k$ elements of $S_{\text{in}}\bigcup S_{\text{out}}$, and so $s_{\text{in},(i)}> s_{\text{out},(k-i+1)}$.
\end{itemize} 


\noindent In our setting, we derive $s_{\text{in}, (i)}$ by noticing that the distribution of the i\up{th} highest value of $S_{\text{in}}$ is also the distribution of its $(k-i)$\up{th} lowest value. Hence, $s_{\text{in}, (i)}$ is the $(k-i)$\up{th} order statistic of $S_{\text{in}}$. As $\mu = 0$, we have that, given two distinct elements of $\mathcal{U}$, say $u_{1}$ and $u_{2}$, $\Cov(s(u_{1}, \mu_{\mathcal{U}}), s(u_{2}, \mu_{\mathcal{U}})) = 0$, which implies that all elements of  $S_{\text{in}}$ are i.i.d. and that the probability density function of the $(k-i)$\up{th} order statistic of $S_{\text{in}}$ is \cite{nagaraja2003order}: 
$$f_{\text{in}, (k-i)}(x) = \frac{k!}{\sigma_{\text{in}}2^{k-1}(k-i)!(i-1)!}\phi(\frac{x-\mu_{\text{in}}}{\sigma_{\text{in}}})(1+\erf(-\frac{x-\mu_{\text{in}}}{\sqrt{2}\sigma_{\text{in}}}))^{k-i}(1-\erf(-\frac{x-\mu_{\text{in}}}{\sqrt{2}\sigma_{\text{in}}}))^{i-1},$$
with:
\begin{align*}
\mu_{\text{in}} &= d\frac{\sigma^{2}}{k}, \\ \sigma_{\text{in}} &= \sigma^{2}\frac{\sqrt{d(\kappa+k-2)}}{k}, \\ \phi &: x \mapsto \frac{2}{\sqrt{\pi}}\int_{0}^{x}e^{-t^2}dt.
\end{align*}
Similarly, $s_{\text{out}, (k-i+1)}$ is the $(N-2k+i)$\up{th} order statistic of $S_{\text{out}}$ and so its cumulative density function is~\cite{nagaraja2003order}:
$$F_{\text{out}, (k-i+1)}(x) = \sum_{j=N-2k+i}^{N-k} \binom{N-k}{j}(1+\erf(-\frac{x}{\sqrt{2}\sigma_{\text{out}}}))^{j}(1-\erf(-\frac{x}{\sqrt{2}\sigma_{\text{out}}}))^{N-k-j},$$
with: 
$$\sigma_{\text{out}} = \sigma^{2}\sqrt{\frac{d}{k}}.$$ Finally, we obtain:
\begin{equation}
\begin{split}
p_{(\frac{i}{k})}^{+} &= \mathbb{P}(s_{\text{in},(i)} > s_{\text{out},(k-i+1)}) \\ 
&= \int_{-\infty}^{\infty}f_{\text{in},(i)}(x)\times F_{\text{out},(k-i+1)}(x)\d x,
\label{eq-proof2-2}
\end{split}
\end{equation}
and, by injecting Equation~\eqref{eq-proof2-2} into Equation~\eqref{eq-proof2-1}, we retrieve the expression from Equation~\eqref{eq:th_result_2} of Proposition~\ref{proposition2}.

\clearpage
\bibliographystyle{ACM-Reference-Format}
\bibliography{references}

\end{document}